\documentclass[twocolumn,floats,floatfix,aps,pra]{revtex4-1}
\usepackage{amsfonts,amssymb,amsmath}
\usepackage{color,calc}
\usepackage[dvips]{graphicx}
\usepackage{bm}

\def\be{ \begin{equation} }
\def\ee{ \end{equation} }
\def\bea{ \begin{eqnarray} }
\def\eea{ \end{eqnarray} }
\def\bse{ \begin{subequations} }
\def\ese{ \end{subequations} }
\def\ba{ \begin{array} }
\def\ea{ \end{array} }

\def\to{\rightarrow}

\def\U{\mathbf{U}}

\def\H{\mathbf{H}}

\newcommand{\ket}[1]{\vert #1\rangle}

\def\phase{\phi}

\def\red{}
\def\black{}

\def\red{\black}

\def\P{\Omega_p}

\def\Re{\textrm{Re}}
\def\Im{\textrm{Im}}
\def\p{p}
\def\q{q}

\def\P{P}
\def\Q{Q}
\def\prob{\mathcal{P}}

\def\PN{P^{(N)}}
\def\QN{Q^{(N)}}
\def\PNc{P^{(N)}_c}
\def\QNc{Q^{(N)}_c}
\def\UN{\U^{(N)}}

\def\PN{P_N}
\def\QN{Q_N}
\def\PNc{P_N^c}
\def\QNc{Q_N^c}
\def\UN{\U_N}

\def\M{M}

\begin{document}


\title{Relations between single-pass and multi-pass qubit gate errors}

\author{Nikolay V. Vitanov}

\affiliation{Department of Physics, St Kliment Ohridski University of Sofia, 5 James Bourchier blvd, 1164 Sofia, Bulgaria}

\date{\today }

\begin{abstract}
In quantum computation the target fidelity of the qubit gates is very high, with the admissible error being in the range from $10^{-3}$ to $10^{-4}$ and even less, depending on the protocol.
The direct experimental determination of such an extremely small error is very challenging \red by standard quantum-process tomography.
Instead, the method of randomized benchmarking, which uses a random sequence of Clifford gates, has become a standard tool for determination of the average gate error as the decay constant in the exponentially decaying fidelity.
In this paper, the task for determining a tiny error is addressed by sequentially repeating the \emph{same} gate multiple times, which leads to the coherent amplification of the error, until it reaches large enough values to be measured reliably.
\black
If the transition probability is $p=1-\epsilon$ with $\epsilon \ll 1$ in the single process, then classical intuition dictates that the probability after $N$ passes should be $\PN \approx 1 - N \epsilon$.
However, this classical expectation is misleading because it neglects interference effects.
This paper presents a rigorous theoretical analysis based on the SU(2) symmetry of the qubit propagator, resulting in explicit analytic relations that link the $N$-pass propagator to the single-pass one \red in terms of Chebyshev polynomials.
In particular, the relations suggest that in some special cases the $N$-pass transition probability degrades as $\PN = 1-N^2\epsilon$, i.e. dramatically faster than the classical probability estimate.
In the general case, however, the relation between the single-pass and $N$-pass propagators is much more involved.
Recipes are proposed for unambiguous determination of the gate errors in the general case, and for both Clifford and non-Clifford gates.
\black
\end{abstract}

\maketitle


\section{Introduction}\label{Sec:intro}
%
In quantum computation the admissible error of gate operations is very small --- usually in the range of $10^{-4}$ to $10^{-3}$. 
Recently, single-qubit gate errors as small as $10^{-5}$ \cite{Brown2011,Gaebler2016} and even $10^{-6}$ \cite{Harty2014}, and two-qubit gate errors below $10^{-3}$ \cite{Ballance2016,Gaebler2016} have been reported in trapped-ions experiments.
The unwanted cross-talk to neighboring qubits has been reduced to $10^{-5}$ \cite{Piltz2014} and $10^{-6}$ \cite{Craik2017} in other trapped-ions experiments,
 and ion transport with an error of less than $10^{-5}$ has been reported too \cite{Kaufmann2018}.
In superconducting qubits, single-qubit gate fidelity of 99.9\% and two-qubit gate fidelity of 99.4\% have been achieved \cite{Barends2014}.

\red
The direct experimental determination of such tiny errors is very challenging with the standard methods of quantum-state and quantum-process tomography.
Alternatively, the recently developed methods of randomized benchmarking and gate set tomography can be used.
Randomized benchmarking characterizes the fidelity of Clifford gates \cite{Emerson2007, Knill2008,Magesan2011,Magesan2012,Wallman2014,Epstein2014, Magesan2012prl,Gambetta2012,Kimmel2014,Wallman2015,Alexander2016,McKay2019} by applying sequences of a large number of $\pi$ pulses in a randomly chosen $x$ or $y$ direction as well as two $\pi/2$ pulses in the beginning and the end of the sequence.
The average gate error is deduced from the decay rate of the fidelity when plotted versus the length of the sequence.
Extensions of randomized benchmarking to non-Clifford gates have been proposed too \cite{Carignan-Dugas2015,Cross2016,Harper2017}.
Gate set tomography \cite{Merkel2013,Blume-Kohout2013,Greenbaum2015,Dehollain2016,Blume-Kohout2017}, on the other hand,
 is a sophisticated recent method which delivers simultaneously the fidelities of a set of gates.


Randomized benchmarking, in particular, is widely used due to its simplicity.
However, it delivers an error value which is not exactly the error of a particular gate but an average error over several gates ($\pi_x$, $\pi_y$, $\pi/2$), which may have different errors.
Recently, there has been some discussion about what randomized benchmarking actually measures \cite{Proctor2017,Carignan-Dugas2018}.

In the quest for simple and fast algorithms for characterizing the gate error here I propose to sequentially repeat the \emph{same} gate many times, and hence amplify the error and make its accurate measurement feasible.
In this manner, because only the characterized gate is used one does not introduce other errors, assuming that the experimental apparatus produces exactly the same gate and the same error sufficiently many times, which is not unreasonable.
Moreover, because randomization is absent, the gate errors add up coherently which implies that the tiny gate error increases much more quickly to measurable values than in a random process.

To this end, it is of crucial importance to have a relation that links the single-pass propagator $\U$ to the $N$-pass propagator $\U_N = \U^N$, and hence the single-pass transition probability $p$ to the $N$-pass transition probability $\PN$.
\black
If the transition probability for the single process is $p=1-\epsilon$ with $0<\epsilon \ll 1$, then classical intuition dictates that the $N$-pass probability should be $\PN = p^N \approx 1 - N \epsilon$.
[A more accurate calculation, which takes into account the exchange of probabilities between the qubit states adds correction terms to $p^N$ (see Sec.~\ref{Sec:classical} below) but in the limit of a tiny error the above estimate for $\PN$ remains in place.]
However, this classical expectation is misleading because it neglects interference effects caused by the dynamical phases in the propagator.

Recently \cite{Vitanov2018}, I analyzed the double-pass transition probability in two-state and three-state quantum systems by using, respectively, the SU(2) and SU(3) symmetry of the corresponding propagator.
The correct double-pass probability is obtained by multiplying the sequential propagators, rather than probabilities.
The conclusion was that the quantum estimate for the error generally exceeds the classical estimate $2 \epsilon$, i.e. the quantum probability degrades faster than the classical one.
I have derived the exact relationship (for any value of $p$) between the single-pass and double-pass probabilities, which in the general case depends on a dynamic phase.
A recipe for the determination of the single-pass probability from the double-pass probability from a pair of two different measurements was proposed.
In two special cases of interest, when the Hamiltonian possesses certain symmetries, the single-pass probability can be determined unambiguously from a single double-pass signal.

These results are of interest in physical situations when it is much easier to measure the initial-state population rather than the target one, e.g. in the formation of ultracold ground-state molecules \cite{Takekoshi2014} from ultracold atoms, and in atomic excitation to Rydberg levels \cite{Higgins2017}.
However, these results are less relevant for the objective mentioned in the beginning --- the determination of a tiny transition probability error $\epsilon$ ($0 < \epsilon \ll 1$) --- because the double-pass error remains still small.

In the present paper, I extend this earlier approach to $N$ repeated processes in a qubit.
I use a rigorous theoretical analysis based on the SU(2) symmetry of the qubit propagator, resulting in an explicit analytic relation that links the $N$-pass propagator to the single-pass one.
\red
In several special cases this relation takes a simple form which allows one to unambiguously deduce the single-pass error from the $N$-pass one.
In the general case, this is not possible but instead, a procedure is proposed which allows one to deduce the single-pass error from several multi-pass signals coming from sequences of interleaved gates.
\black

\section{Classical probability}\label{Sec:classical}
Assuming that initially the system is in state $\ket{1}$ let us denote by $\QN= \prob_{1\to 1}^{(N)}$ and $\PN = \prob_{1\to 2}^{(N)}$ the probabilities for, respectively, return to state $\ket{1}$ and transition to state $\ket{2}$.
The exact classical probabilities after $N$ passes are
\be\label{classical}
\QNc = \frac{1 + (1-2p)^N}{2} , \quad
\PNc = \frac{1 - (1-2p)^N}{2} ,
\ee
where $p$ is the single-pass transition probability.
These formulas can easily be proved inductively by using the relations $Q_{N+1}^c = (1-p) \QNc + p \PNc$ and $P_{N+1}^c = p \QNc + (1-p) \PNc$.

(i) For large probability $p=1-\epsilon$ ($0 < \epsilon \ll 1$), we have

\begin{itemize}
\item $\QNc \approx 1 - N\epsilon $ and $\PNc \approx N\epsilon $ for even $N$;

\item $\QNc \approx N\epsilon $ and $\PNc \approx 1 - N\epsilon $ for odd $N$.
\end{itemize}

In this limit, this is the same behavior as the simple estimates $p^N$ and $1-p^N$, which neglect the mutual exchange of probabilities between the two states: the error increases linearly with the number of processes $N$.

(ii) For small probability $p=\epsilon$ ($0 < \epsilon \ll 1$), we have
$\QNc \approx 1 - N\epsilon $ and $\PNc \approx N\epsilon $ for any $N$ (odd or even).
Hence the transition probability grows as $N$.

\red

(iii) For half probability $p=\frac12-\epsilon$ ($0 < \epsilon \ll 1$), we have $\QNc = \frac12 + \frac12 (2\epsilon)^N$ and $\PNc = \frac12 - \frac12 (2\epsilon)^N$.
Hence in this case the error rapidly vanishes, rather than increases, with $N$ and the probabilities tend to $\frac12$.

(iv) For intermediate probability $p=p_0 -\epsilon$ ($0 < \epsilon \ll 1$), we find $\QNc \approx \frac12 [1+(1-2p_0)^N] + N (1-2p_0)^{N-1} \epsilon$ and $\PNc = 1-\QNc$.
Here again the error rapidly vanishes because $N (1-2p_0)^{N-1} \to 0$ as $N$ increases  (for $p_0\neq 0,1$), and the probabilities tend to $\frac12$.

\black

Below the quantum probabilities after $N$ sequential passes are derived, discussed and compared to the classical probabilities.


\section{Multi-pass probabilities}\label{Sec:multi-pass}
\begin{figure}[t]
\includegraphics[width=0.95\columnwidth]{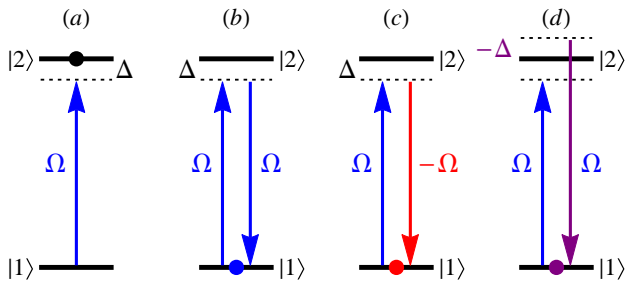}
\caption{
Set of measurements required to measure the single-pass transition probability (a) by multiple applications of the same process, or by multiple applications of several double-pass processes: (b) with the same Rabi frequencies, (c) with different signs of the Rabi frequencies, (d) with different signs of the detunings.
}
\label{fig-2ss}
\end{figure}

%
The Hamiltonian of a coherently driven lossless two-state quantum system, in the rotating-wave approximations \cite{Shore1990}, reads
\be\label{H-2}
\H(t) = \tfrac12 \left[\begin{array}{cc} -\Delta(t) & \Omega(t) \\ \Omega(t) & \Delta(t) \end{array}\right],
\ee
where $\Delta(t)$ is the system-field frequency mismatch (the detuning), and $\Omega(t)$ is the Rabi frequency, which is a measure of the coupling between the two states.
For arbitrary $\Omega(t)$ and $\Delta(t)$ the corresponding propagator is a SU(2) matrix, which can be expressed in terms of the complex-valued Cayley-Klein parameters $a$ and $b$ ($|a|^2 +|b|^2=1$) as
\be\label{U-2}
\U = \left[\begin{array}{cc} a & -b^* \\ b & a^* \end{array}\right].
\ee
Assume that the system is initially in state $\ket{1}$.
Then the probabilities for remaining in state $\ket{1}$ and for transfer to state $\ket{2}$ are
\be\label{2ss-pq}
\q = \prob_{1\to 1} = |a|^2,\quad \p = \prob_{1\to 2} = |b|^2. 
\ee
Obviously, $p+q = 1$.
If the transition probability is very close to 1, i.e. $\p = 1- \epsilon$, with $0 < \epsilon \ll 1$, then it is difficult to determine the error $\epsilon$ precisely.
A natural approach that is adopted here is to repeat the process $N$ times, which amplifies the error, see Fig.~\ref{fig-2ss}.
By measuring the population of state $\ket{1}$ or $\ket{2}$ after $N$ passes one can \emph{deduce} the single-pass transition probability $p$, and hence the single-pass error $\epsilon$.

In order to determine the populations after $N$ passes we need to find the $N$-pass propagator $\UN = \U^{N}$.
It has been proved \cite{Vitanov1995} that the $N$-th power of any SU(2) propagator, parameterized as in Eq.~\eqref{U-2}, reads
\be\label{U-N}
\UN = \left[\begin{array}{cc}
\cos N\theta + ia_i \dfrac{\sin N\theta}{\sin \theta} & -b^* \dfrac{\sin N\theta}{\sin \theta} \\
 b \dfrac{\sin N\theta}{\sin \theta} &  \cos N\theta  - ia_i \dfrac{\sin N\theta}{\sin \theta}
\end{array}\right],
\ee
where $a = a_r + i a_i$ and
\be\label{theta}
\theta = \arccos a_r \quad (0 \leqq \theta \leqq \pi).
\ee
Therefore, the two populations after $N$ passes are
\bse\label{PN,QN}
\begin{align}
\QN &= \prob_{1\to 1}^{(N)} = 1 - p \frac{\sin^2 N\theta}{\sin^2 \theta}, \label{QN} \\
\PN &= \prob_{1\to 2}^{(N)} = p \frac{\sin^2 N\theta}{\sin^2 \theta}. \label{PN}
\end{align}
\ese

In the general case, measuring $\QN$ or $\PN$ alone is not sufficient to deduce the single-pass transition probability $p$ because the parameter $\theta$ is not uniquely linked to $p$ but it depends also on the phase of the propagator parameter $a$.
However, if $a$ is real, then this is possible.

\begin{figure}[t]
\begin{tabular}{c}
\includegraphics[width=0.90\columnwidth]{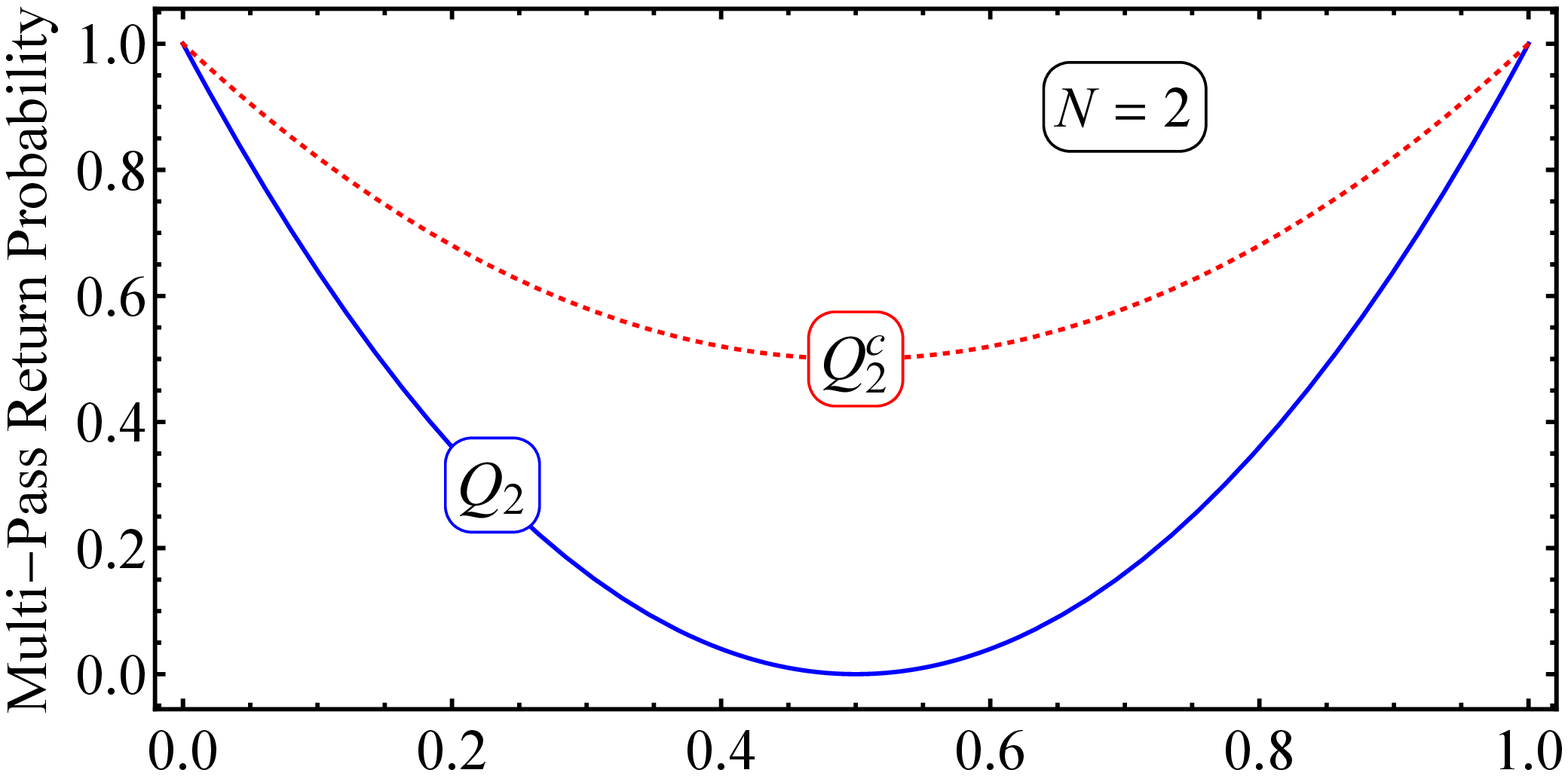} \\
\includegraphics[width=0.90\columnwidth]{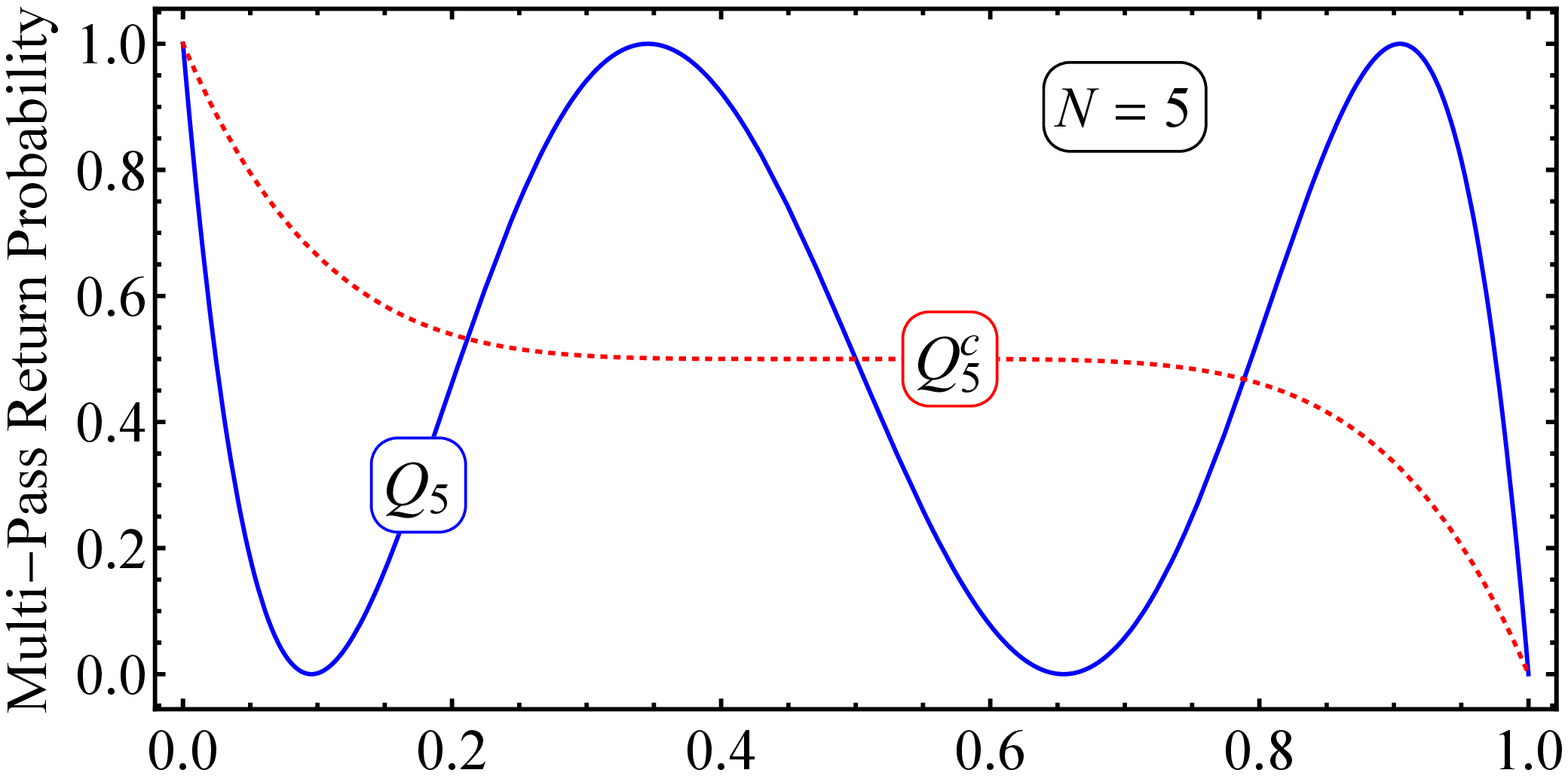} \\
\includegraphics[width=0.90\columnwidth]{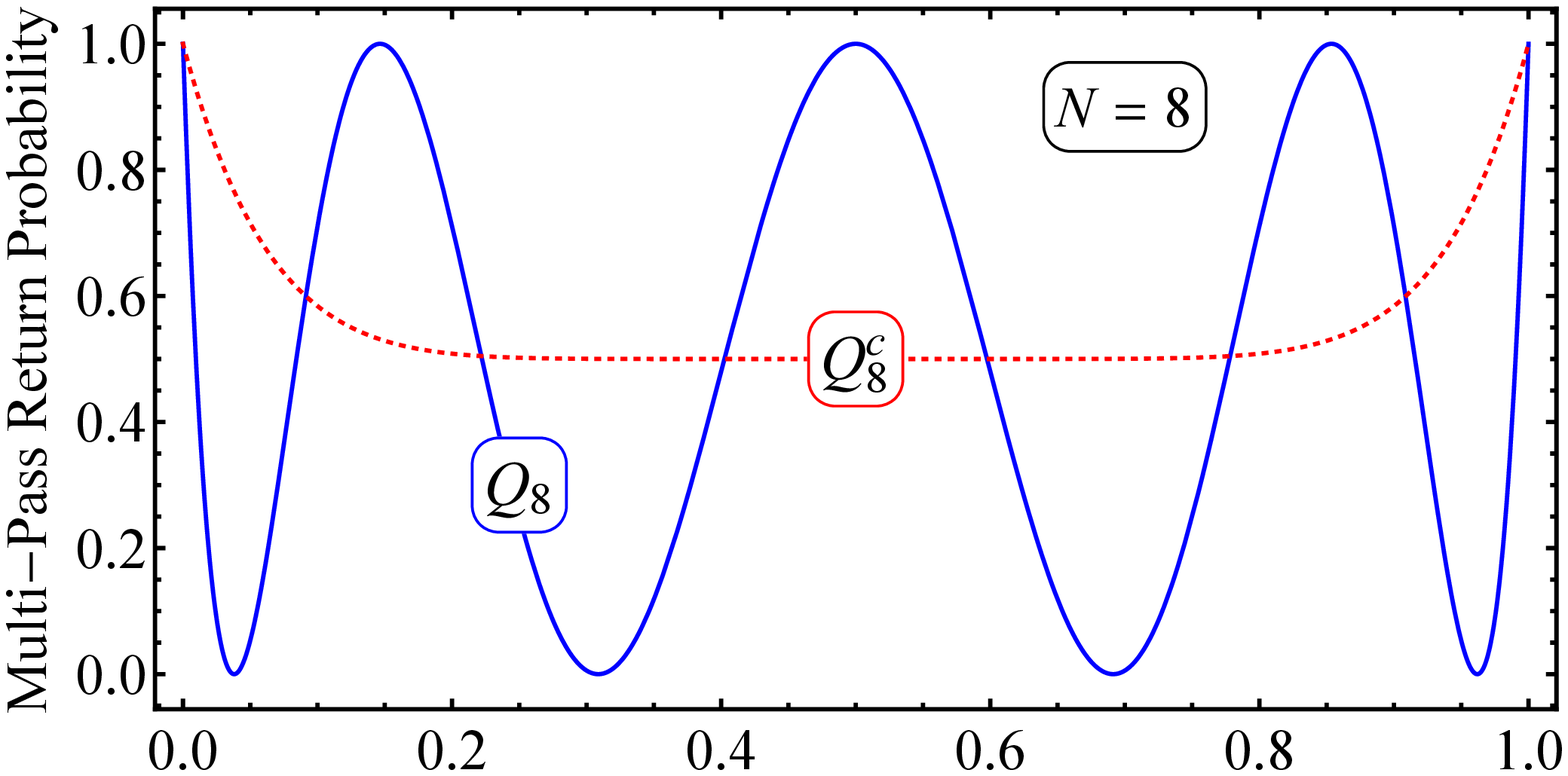} \\
\includegraphics[width=0.90\columnwidth]{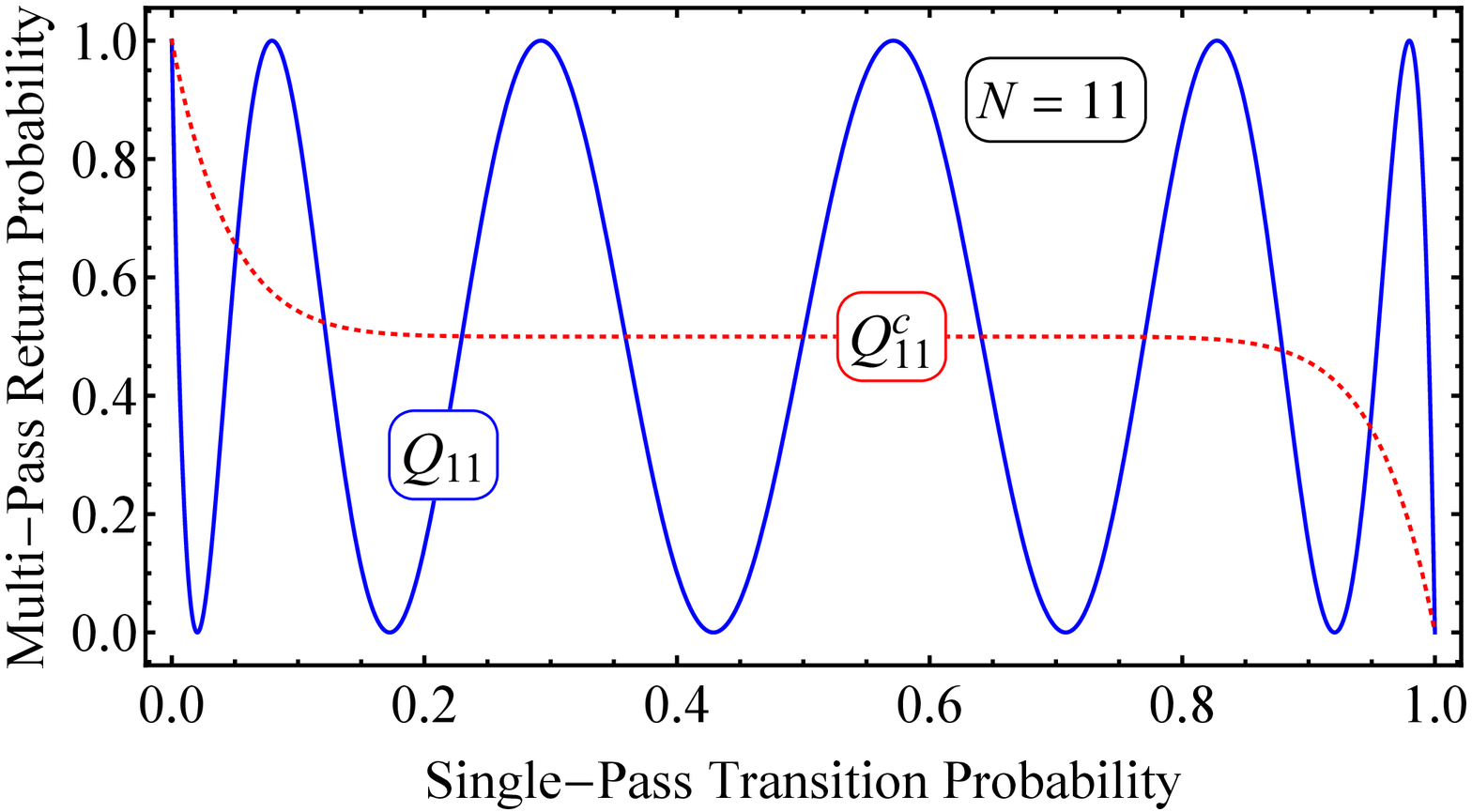}
\end{tabular}
\caption{
Multi-pass probability $\QN$ for return to the initial state vs the single-pass transition probability $p$ for $N=2$, 5, 8, and 11 passes for a real-valued Cayley-Klein parameter $a$.
In each frame the solid curve is the quantum probability of Eq.~\eqref{Q_real} and the dashed curve is the classical probability $\QNc$ of Eq.~\eqref{classical}.
}
\label{fig:pQ}
\end{figure}

\section{Special case: Real $a$}\label{Sec:real}

\subsection{Error amplification}\label{Sec:real_error}

\begin{figure}[t]
\begin{tabular}{c}
\includegraphics[width=0.90\columnwidth]{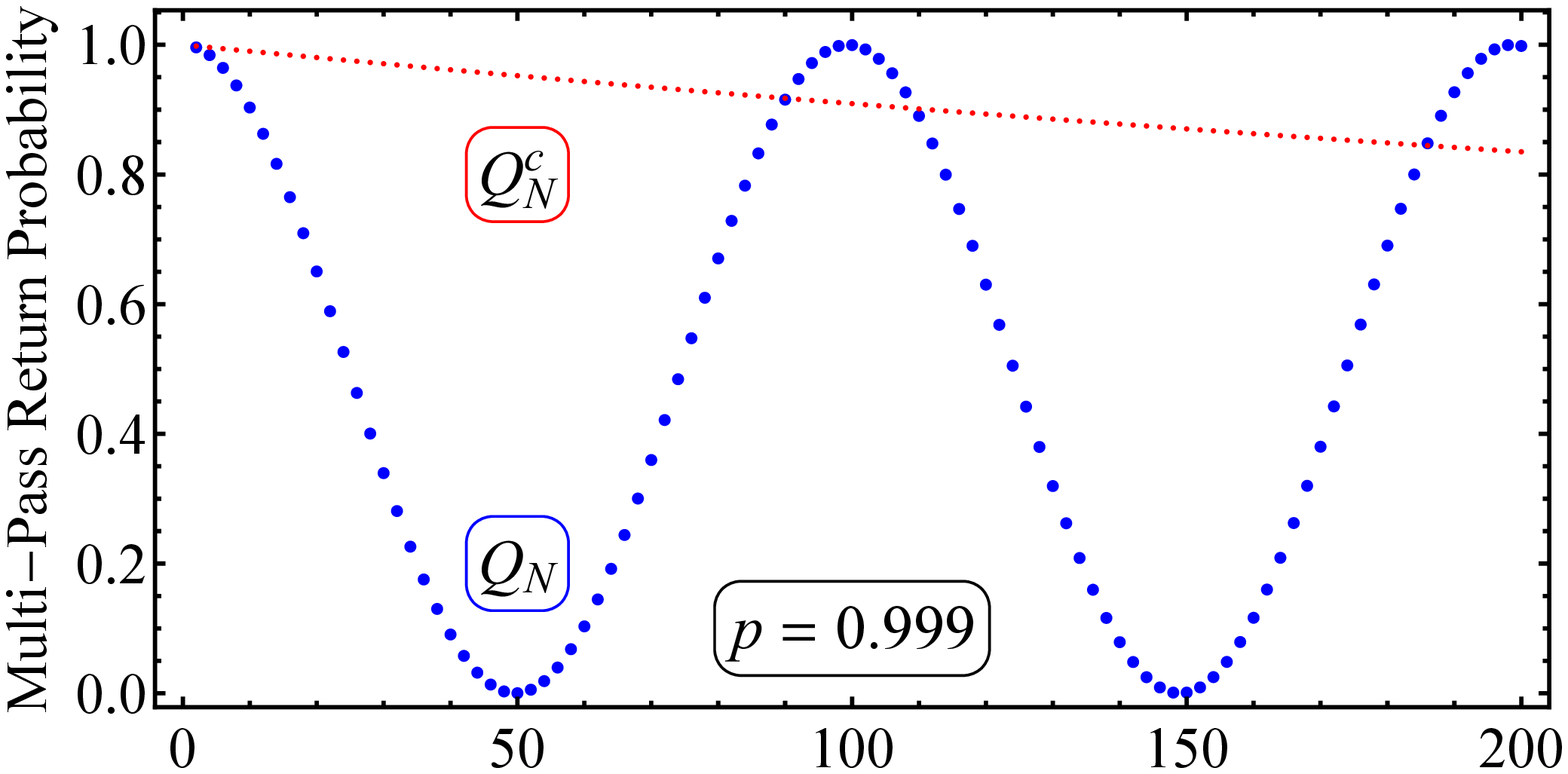} \\
\includegraphics[width=0.90\columnwidth]{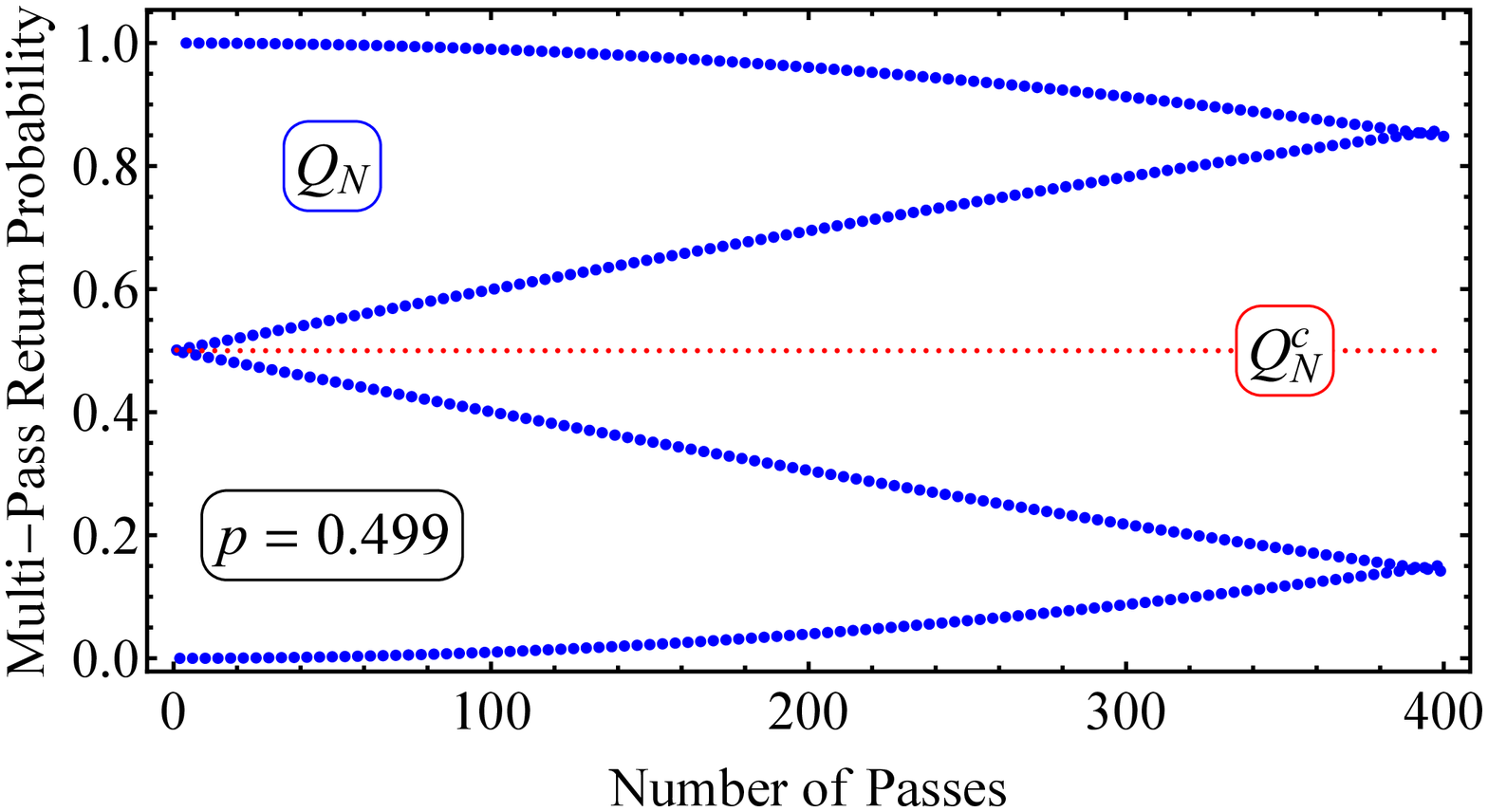}
\end{tabular}
\caption{ {
Multi-pass probability $\QN$ for return to the initial state vs number of passes $N$ for single-pass transition probabilities $p=0.999$ (top) and 0.499 (bottom) and a real-valued Cayley-Klein parameter $a$.
In each frame the dotted curve is the quantum probability of Eq.~\eqref{Q_real} and the dashed curve is the classical probability $\QNc$ of Eq.~\eqref{classical}.
In the top frame only the dots for even $N$ are plotted.
In the bottom frame the four branches for $\QN$ correspond to the four expressions of Eqs.~\eqref{Q-half}.}
}
\label{fig:pQ-N}
\end{figure}

For real $a$, we have $q = a^2 = \cos^2\theta$ and $p = \sin^2\theta = 1-q$.
The $N$-pass probabilities \eqref{PN,QN} become
\bse\label{QP_real_a}
\begin{align}
\QN &= \cos^2 [N \arccos(q^{\frac12})] = T_N(q^{\frac12}), \label{Q_real}\\
\PN &= \sin^2 [N \arccos(q^{\frac12})] = 1-T_N(q^{\frac12}), \label{P_real}
\end{align}
\ese
where $T_{N}(x)$ denotes the Chebyshev polynomial of the first kind.
The return probability $\QN$ is plotted in Fig.~\ref{fig:pQ} versus the single-pass transition probability $p$ and compared to the classical return probability $\QNc$ of Eq.~\eqref{classical}.
\red
The return probability $\QN$ is also plotted in Fig.~\ref{fig:pQ-N} versus the number of passes $N$.
The discrepancy between the classical and quantum probabilities in all cases is drastic.
Note that for any single-pass probability $p$ the multi-pass probabilities are periodic functions of $N$, as evident in Fig.~\ref{fig:pQ-N} (top), and there exists an optimum range of passes for which the error is maximized.

\red

Several important special cases follow.

(i) For $p=1-\epsilon$ $(\epsilon \ll 1)$ we find the return probability to be
\bse\label{P_real_a}
\begin{align}
\QN &\approx 1 - N^2 \epsilon + O(\epsilon^2) \quad (\text{even}~N), \label{Q_real_a_even} \\
\QN &\approx N^2 \epsilon + O(\epsilon^2) \quad (\text{odd}~N), \label{Q_real_a_odd}
\end{align}
\ese
while the classical probability is $\QNc \approx N \epsilon$ for odd $N$ and $\QNc \approx 1 - N \epsilon$ for even $N$.
Obviously, the quantum probability error increases much faster (as $N^2$) with the number of passes than the classical probability (as $N$).
This is clearly visible near $p=1$ in all frames of Fig.~\ref{fig:pQ} and in Fig.~\ref{fig:pQ-N} (top).
Error amplification to probability values of about $\frac12$ occurs for $N = \lfloor1/\sqrt{2\epsilon}\rfloor$ passes.

(ii) In the opposite limit, when $p =\epsilon \ll 1$ is small, we find
\bse\label{QP_real_a_small}
\begin{align}
\QN &\approx 1 - N^2 \epsilon + O(\epsilon^2)\quad (\epsilon\ll 1), \label{Q_real_a_small} \\
\PN &\approx N^2 \epsilon + O(\epsilon^2)\quad (\epsilon\ll 1), \label{P_real_a_small}
\end{align}
\ese
i.e. the $N$-pass transition probability grows with $N^2$.
The classical probability again behaves linearly, $\PNc \approx N \epsilon$.
This is clearly visible near $p=0$ in all frames of Fig.~\ref{fig:pQ}.
Again, as for $p=1-\epsilon$, error amplification to probability values of about $\frac12$ occurs for $N = \lfloor1/\sqrt{2\epsilon}\rfloor$ passes.

(iii) In another important special case, when $p=\frac12 -\epsilon$ we find the transition probability to be $ (|\epsilon| \ll 1)$:
\bse\label{Q-half}
\begin{align}
\QN &\approx 1 - N^2 \epsilon^2 + O(\epsilon^4) \quad (N=4k); \label{Q-half-4k} \\
\QN &\approx \tfrac12 + N \epsilon + O(\epsilon^2) \quad (N=4k+1), \label{Q-half-4k+1}\\
\QN &\approx N^2 \epsilon^2 + O(\epsilon^4) \quad (N=4k+2), \label{Q-half-4k+2}\\
\QN &\approx \tfrac12 - N \epsilon + O(\epsilon^2) \quad (N=4k+3), \label{Q-half-4k+3}
\end{align}
\ese
In this case, the $N^2\epsilon$ scaling of the error is replaced by either $N\epsilon$ or $(N\epsilon)^2$.
This is slower than in the previous two cases, even for the $(N\epsilon)^2$ cases because the $N^2$ scaling is negated by the factor $\epsilon^2$ ($ \ll \epsilon$).
Yet, this is very different from the classical case (Sec.~\ref{Sec:classical}) in which, instead of increasing, the error rapidly vanishes with $N$.

The four branches of Eqs.~\eqref{Q-half} describe distinctly different values of the return probability, which are illustrated in Fig.~\ref{fig:pQ-N} (bottom) for $p=\frac12-\epsilon$ with $\epsilon=10^{-3}$.
The two middle branches emanating from the value $\frac12$ and corresponding to Eqs.~\eqref{Q-half-4k+1} and \eqref{Q-half-4k+3} diverge linearly from this value.
The two outer branches emanating from the values 0 and 1 and corresponding to Eqs.~\eqref{Q-half-4k} and \eqref{Q-half-4k+2} diverge quadratically versus $N$, but with the squared error $\epsilon^2=10^{-6}$.
Hence initially the middle branches amplify the error faster until eventually the outer branches catch up, which takes place at about $(\sqrt{3}-1)/(2\epsilon)$ [see Eqs.~\eqref{Q-half-4k} and \eqref{Q-half-4k+1}].
Therefore, for the purpose of error amplification in shortest time the linear middle branches are the suitable ones, with the number of passes required of the order of $1/(4\epsilon)$.


\black

\subsection{Implementations}\label{Sec:real_implementations}

The Cayley-Klein parameter $a$ is real in two important special cases.

(i) On exact resonance ($\Delta=0$) we have $a = \cos (A/2)$, with $A$ being the pulse area.
Hence $\theta = A/2$ and therefore, after $N$ passes we find from Eq.~\eqref{PN}
\be\label{P-small}
\PN = \sin^2 (NA/2).
\ee
Of course, this result can be found directly from the resonant solution because the total pulse area after $N$ passes is $NA$.

(ii) When the Rabi frequency is symmetric and the detuning is anti-symmetric function of time, $\Omega(-t) = \Omega(t)$ and $\Delta(-t) = -\Delta(t)$ the parameter $a$ is real \cite{Vitanov1999nonlinear}.
A number of analytically soluble models belong to this special class: the original Landau-Zener-St\"uckelberg-Majorana (LZSM) model \cite{Landau1932,Zener1932,Stueckelberg1932,Majorana1932}, the symmetric finite LZSM model \cite{Vitanov1996}, the Allen-Eberly-Hioe model \cite{Allen1975,Hioe1984}, and the linearly-chirped Gaussian model \cite{Vasilev2005} --- all related to the popular technique of rapid adiabatic passage (RAP) via a level crossing \cite{Vitanov2001}.

In either of these cases (i) and (ii), the multi-pass probabilities $\PN$ and $\QN$ depend on the single-pass transition probability $p$ only and do not depend on the propagator phases.
Hence the mappings $p\to \PN$ or $p\to \QN$ are single-valued: knowing $p$ means knowing the multi-pass probabilities.

However, the opposite correspondences $\PN \to p$ and $\QN \to p$ are not single-valued (because the $\arccos$ function is not single-valued), as can be deduced also from Fig.~\ref{fig:pQ}.
Therefore, if one wants to find the single-pass probability $p$ from the multi-pass ones $\PN$ or $\QN$, some additional knowledge is required.
For instance, if the probability  $p$ is known (or measured) to be close to 1 then the largest value for $p$ stemming from the value of $\PN$ should be retained.
Alternatively, one can determine $p$ from two measurements, e.g. after $N$ and $2N$ passes, which should produce the same value for $p$.

\section{General case}\label{Sec:general}

\red

In the general case when the parameter $a$ is not real it is not possible to determine the single-pass error by repeating the same interaction $N$ times.
The reason is that $\theta$ is not uniquely linked to $p$ but it depends also on a dynamical phase, see Eq.~\eqref{theta}.
This undesired dependence can be eliminated by using sequences of interleaved pairs of gates (i.e. double-pass processes, see Fig.~\ref{fig-2ss}), in which the second gate is different from, but related to the first one, similarly to Ref.~\cite{Vitanov2018}.
This section discusses this general case.
\black

\subsection{Double-pass propagators}

Consider a second interaction with the same magnitudes but with different signs of $\Omega(t)$ and $\Delta(t)$, cf.~Fig.~\ref{fig-2ss}.
The respective propagators can be obtained from Eq.~\eqref{U-2} by simple algebraic operations \cite{Vitanov1999nonlinear} and, very importantly, can be expressed with the \emph{same} Cayley-Klein parameters $a$ and $b$,
\be\label{2ss-U}
\U_{-\Omega,\Delta} = \left[\begin{array}{cc} a & b^* \\ -b & a^* \end{array}\right], \quad
\U_{\Omega,-\Delta} = \left[\begin{array}{cc} a^* & b \\ -b^* & a \end{array}\right].
\ee
The respective double-pass propagators read
\bse\label{2ss-UU}
\begin{align}
\U_{\Omega,\Delta} \U_{\Omega,\Delta} &= \left[\begin{array}{cc} a^2 - |b|^2 & -2 b^* \Re(a) \\ 2 b \Re(a) & (a^*)^2 - |b|^2 \end{array}\right], \label{2ss-UppU} \\
\U_{-\Omega,\Delta} \U_{\Omega,\Delta} &= \left[\begin{array}{cc} a^2 + |b|^2 & -2 i b^* \Im(a) \\ -2 i b \Im(a) & (a^*)^2 + |b|^2 \end{array}\right], \label{2ss-UmpU} \\
\U_{\Omega,-\Delta} \U_{\Omega,\Delta} &= \left[\begin{array}{cc} |a|^2 + b^2 & 2 i a^* \Im(b) \\ 2 i a \Im(b) & |a|^2 + (b^*)^2 \end{array}\right], \label{2ss-UpmU} \\
\red
\U_{-\Omega,-\Delta} \U_{\Omega,\Delta} &= \left[\begin{array}{cc} |a|^2 - b^2 & -2 a^* \Re(b) \\ 2 a \Re(b) & |a|^2 - (b^*)^2 \end{array}\right]. \label{2ss-UmmU}
\black
\end{align}
\ese
where $\U_{\Omega,\Delta}$ is the same as $\U$ of Eq.~\eqref{U-2} but the subscripts are added for the sake of consistency with the other propagators.
We immediately see that the corresponding double-pass probabilities are different in each case; it was this difference that enabled the derivation of the single-pass probability $p$ from the double-pass probabilities in Ref.~\cite{Vitanov2018}.

For $\M$ double-pass pairs of this type, i.e. for $2\M$ passes, we can find the overall propagator in the same manner as Eq.~\eqref{U-N} is derived from Eq.~\eqref{U-2}.

\subsection{Special case: Imaginary $b$}
Imaginary $b$ occurs when both the Rabi frequency and the detuning are symmetric functions of time, $\Omega(-t) = \Omega(t)$ and $\Delta(-t) = \Delta(t)$ \cite{Delos1972,Bambini1984,Vitanov1999nonlinear}.
A beautiful analytically soluble model that belongs to this class is the Rosen-Zener model \cite{Rosen1932}, in which the Rabi frequency has a hyperbolic-secant shape, $\Omega(t) \propto \textrm{sech} (t/T)$ and the detuning is constant, $\Delta=$\,const.
Another (approximately soluble) example is the Gaussian model with constant detuning \cite{Vasilev2004}, in which $\Omega(t) \propto e^{-t^2/T^2}$.
In these cases, the Cayley-Klein parameter $b$ is purely imaginary \cite{Delos1972,Bambini1984,Vitanov1999nonlinear}, implying $p = |b|^2 = [\Im(b)]^2$.
Then the diagonal elements of the product $\U_{\Omega,-\Delta} \U_{\Omega,\Delta}$ in Eq.~\eqref{2ss-UpmU} are real and we find $\theta = \arccos(1-2p)$.
Therefore $\sin^2 (\theta) = 4p(1-p)$ and hence
\be
P_{2\M} = \sin^2 (\M\theta) = 1 - T_{\M}(1-2p),
\ee
which is the same as in the previous case, Eq.~\eqref{QP_real_a}, with the replacement $N = 2\M$ (because here we have $\M$ double-pass processes)
\red
 and noting that $T_{2\M}(q^\frac12) = T_{\M}(2q-1) = T_{\M}(1-2p)$.

For large and small transition probability, $p=1-\epsilon$ and $p=\epsilon$ with $0<\epsilon \ll 1$, we find
\bse\label{Q-Im}
\begin{align}
\Q_{2\M} &\approx 1 - 4\M^2 \epsilon + O(\epsilon^2) \quad (p = 1-\epsilon), \label{Q-Im-1} \\
\Q_{2\M} &\approx 1 - 4\M^2 \epsilon + O(\epsilon^2) \quad (p = \epsilon). \label{Q-Im-0}
\end{align}
\ese
These are the same expressions as Eq.~\eqref{Q_real_a_even} and Eq.~\eqref{Q_real_a_small}, under the replacement $N = 2\M$, with the same quadratic scaling of the probability error with the number of passes $\M$.

 \black

\subsection{General case: Complex $a$ and $b$}\label{Sec:general}
When the Cayley-Klein parameter $a$ is complex, $a = \sqrt{q} e^{i\xi}$ the parameter $\theta$ cannot be linked to the single-pass transition probability $p=1-q$ alone.
Instead we have $\cos\theta = \sqrt{q} \cos\xi$, i.e. $\theta$ depends on the phase $\xi$ (known also as a St\"uckelberg phase).
Therefore we cannot use the approaches in the two special cases of real $a$ or imaginary $b$ described above.
One possibility to proceed is to extend the approach of Ref.~\cite{Vitanov2018}, which uses the double-pass propagators $\U_{\Omega,\Delta} \U_{\Omega,\Delta}$ and $\U_{-\Omega,\Delta} \U_{\Omega,\Delta}$, see Eqs.~\eqref{2ss-UppU} and \eqref{2ss-UmpU}.
For these double-pass propagators the parameter $\theta$ is defined as
\be\label{theta+-}
\theta_\pm = \arccos (q \cos2\xi \mp p),
\ee
where $\theta_\pm$ refers to $\U_{\pm\Omega,\Delta} \U_{\Omega,\Delta}$, and the respective two-pass transition probabilities are
  $p_+ = 4p(1-p)\cos^2 \xi $ and $p_- = 4p(1-p)\sin^2 \xi $. 
Hence the transition probability after $\M$ double passes depends on both $p$ (or $q$) and $\xi$, cf. Eq.~\eqref{PN},
\red
\bse\label{PN-general}
\begin{align}
\P_{2\M}^{++} &= p_+ \frac{\sin^2 (\M \theta_+)}{\sin^2 (\theta_+)}, \\
\P_{2\M}^{-+} &= p_- \frac{\sin^2 (\M \theta_-)}{\sin^2 (\theta_-)}.
\end{align}
\ese
Here $\P_{2\M}^{++}$ and $\P_{2\M}^{-+}$ are the transition probabilities after $\M$ repetitions of the double-pass propagators $\U_{\Omega,\Delta} \U_{\Omega,\Delta}$ [Eq.~\eqref{2ss-UppU}] and $\U_{-\Omega,\Delta} \U_{\Omega,\Delta}$ [Eq.~\eqref{2ss-UmpU}], respectively.


For $p = 1 - \epsilon$ with $0 < \epsilon \ll 1$, we find from Eq.~\eqref{PN-general}
\bse
\begin{align}
\P_{2\M}^{++} &= 4\M^2 \epsilon \cos^2\xi + O(\epsilon^2), \\
\P_{2\M}^{-+} &=  4\M^2 \epsilon \sin^2\xi + O(\epsilon^2),
\end{align}
\ese
and hence, the multi-pass transition probability error scales as $\M^2$, as before, although now it depends also on the dynamical phase $\xi$ too.
Obviously, the dependence on $\xi$ can be eliminated in the first order of $\epsilon$ by taking the sum $\P_{2\M}^{++} + \P_{2\M}^{-+} = 4\M^2 \epsilon + O(\epsilon^2)$, and one can determine the single-pass error $\epsilon$ from here as
\be\label{error-P=1-xi}
\epsilon \approx \frac{ \P_{2\M}^{++} + \P_{2\M}^{-+} }{4\M^2} .
\ee
However, the higher orders of $\epsilon$ still depend on $\xi$ and therefore, Eq.~\eqref{error-P=1-xi} can be used for sufficiently small values of $\M^2 \epsilon$ only.

For $p=\frac12-\epsilon$, a simple protocol is not easy to find and one can follow the general measurement procedure outlined below.
However, it should be pointed out that quantum gates of this type, e.g. the Hadamard gate, are usually produced by a resonant $\pi/2$ pulse, for which the Cayley-Klein parameter $a$ is real and one can deduce the gate error by using the simple method of Sec.~\ref{Sec:real}.
Nevertheless, if the parameter $a$ acquires a small nonzero phase $\xi$, the parameter $\theta = \arccos(\sqrt{q} \cos\xi)$ will depend very weakly on $\xi$ [as $O(\xi^2)$] and its effect may be negligible.


\black

For general values of $p$, we need more steps.
A possible scenario is the following.
First, measure the transition probability after $\M$ and $2\M$ double passes and calculate the ratios
\be
R_{\M}^\pm = \frac{ P^{\pm +}_{4\M} }{ \P_{2\M}^{\pm +} } = \frac{\sin^2 (2\M\theta_\pm) }{\sin^2 (\M\theta_\pm)} = 4 \cos^2 (\M\theta_\pm).
\ee
In this manner the probabilities $p_\pm$ 
 are eliminated and each of the ratios $R_{\M}^\pm$ depends on the single parameter $\theta_\pm$ only (rather than on $\theta_\pm$ \emph{and} $p_\pm$, as do $\P_{2\M}^{\pm +}$).
From here we find
\be
\theta_\pm = \frac{\arccos(R_{\M}^\pm/2 - 1)}{2\M} ,
\ee
and hence we can find $p$ from [cf.~Eq.~\eqref{theta+-}]
\be
p = \frac{\cos\theta_- - \cos\theta_+} {2}.
\ee
This procedure allows one to determine any single-pass transition probability: small, large or in between.

\red

For $p=\epsilon \ll 1$, the error amplification is more difficult to achieve in the general case.
 It is considered in more detail in the next section, because this case refers to the very important quantum phase gate.

\black

\red
\subsection{Quantum phase gate\label{Sec:phase_gate}}

\def\phaseerror{\gamma}
\def\phase{\alpha}

As a non-Clifford gate, the single-qubit phase gate $\mathbf{F} = e^{i\phase \sigma_z}$ is not amenable to the standard randomized benchmarking, but to its extensions \cite{Carignan-Dugas2015,Cross2016,Harper2017}.
It can be characterized also by the present error amplification protocol.
The single-qubit phase gate can be characterized by sandwiching it between two Hadamard gates, which convert it into a rotational gate.
The the phase gate error is mapped onto the rotational gate error; however, the Hadamard gates will introduce their own errors.
In principle, both the errors of the Hadamard gate and the rotational gate can be amplified and measured using the protocol described in the preceding sections.
However, the multi-pass error amplification method can be used to directly determine the phase gate errors, without Hadamard gates.

In the high-fidelity regime, the off-diagonal elements are very small (meaning transition probability $p =\epsilon \ll 1$), and the target phase $\alpha$ may have its own error $\phaseerror$.
In matrix form the (imperfect) phase gate can be written as
\be\label{phase gate}
\mathbf{F} = \left[\begin{array}{cc} e^{i \xi} \sqrt{1-p} &  e^{i\eta} \sqrt{p}  \\ -e^{-i\eta} \sqrt{p} & e^{-i \xi} \sqrt{1-p} \end{array}\right] ,
\ee
where $\xi = \phase + \phaseerror$, while $\eta$ is the off-diagonal element phase.
The important error terms here are $p$ and $\phaseerror$, both assumed small.
They can be amplified by multiple double passes and determined as follows.

\begin{figure}[tb]
\begin{tabular}{c}
\includegraphics[width=0.90\columnwidth]{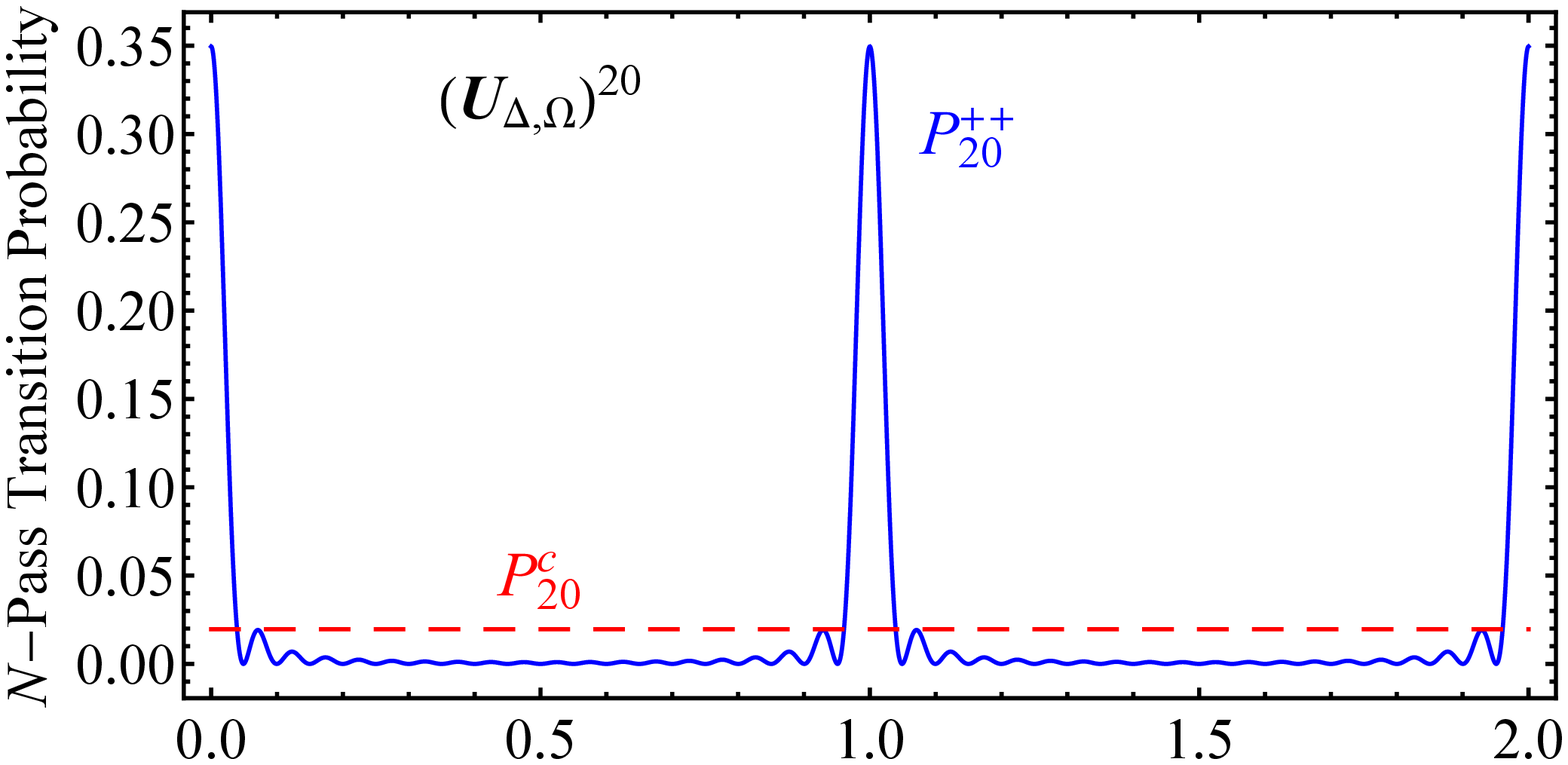}\\
\includegraphics[width=0.90\columnwidth]{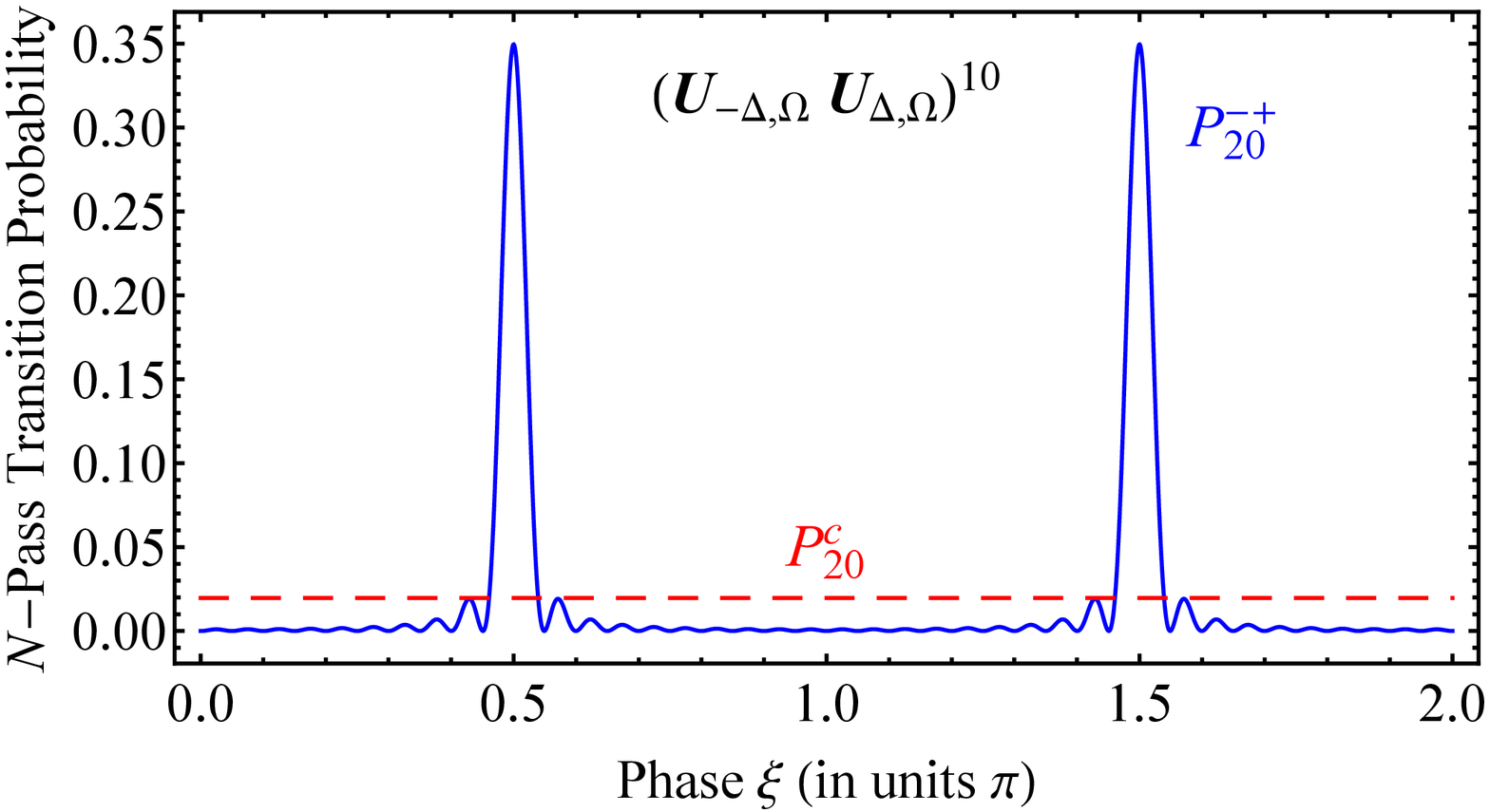}
\end{tabular}
\caption{
Multi-pass transition probabilities $P_{2\M}^{++}$ (top) and  $P_{2\M}^{-+}$ (bottom) of Eqs.~\eqref{PN-general} vs the phase $\xi$ of the Cayley-Klein parameter $a$ for $\M=10$ double passes (hence $N=20$ passes) and single-pass transition probability $p=0.001$.
The solid curves show the exact probabilities 
 and the dashed lines ($P_{20}^c$) show the classical value \eqref{classical}.
}
\label{fig:small}
\end{figure}

%
When 
 $p = \epsilon$ ($0<\epsilon \ll 1$), we find from Eqs.~\eqref{2ss-UU} and \eqref{PN-general} that
\bse\label{PN-small-approx}
\begin{align}
P_{2\M}^{++} &\approx \frac{\sin^2(2\M \xi)}{\sin^2(\xi)} \epsilon + O(\epsilon^2), \label{PN-small-approx-pp} \\
P_{2\M}^{-+} &\approx \frac{\sin^2(2\M \xi)}{\cos^2(\xi)} \epsilon + O(\epsilon^2), \label{PN-small-approx-mp} \\
P_{2\M}^{+-} &\approx 4\M^2 \sin^2(\eta) \epsilon + O(\epsilon^2), \label{PN-small-approx-pm} \\
P_{2\M}^{--} &\approx 4\M^2 \cos^2(\eta) \epsilon + O(\epsilon^2). \label{PN-small-approx-mm}
\end{align}
\ese
These four expressions give the transition probabilities after $\M$ double passes corresponding to the four double-pass propagators of Eqs.~\eqref{2ss-UU}, respectively.
The first expression \eqref{PN-small-approx-pp}, which results from the repetition of the same gate $2\M$ times, resembles the formula for the light intensity distribution of a diffraction grating \cite{Vitanov1995}.
Figure \ref{fig:small} (top) illustrates this dependence.
Obviously, this probability depends very strongly on the phase $\xi$.
The transition probability is suppressed well bellow the classical value \eqref{classical} for most of the range, with the exception of the ranges around $\xi = 0,\pi,2\pi$, where it has the approximate value $4\M^2 \epsilon$.

Obviously, the multi-pass error amplification does not occur in the probabilities, except for $\xi \approx 0,\pi,2\pi$ for $P_{2\M}^{++}$ [Fig.~\ref{fig:small} (top)]
 and $\xi \approx \pm \pi/2$ for $P_{2\M}^{-+}$ [Fig.~\ref{fig:small} (bottom)], which are therefore not suitable for characterization of a general phase gate.
However, the probabilities \eqref{PN-small-approx-pm} and \eqref{PN-small-approx-mm} obtained with interleaved gates do offer error amplification, despite the emergence of the unwanted St\"uckelberg phase $\eta$.
This phase can be eliminated by taking the sum
\be
P_{2\M}^{+-} + P_{2\M}^{--} \approx 4\M^2 \epsilon + O(\epsilon^2), \label{PN-small-approx-sum}
\ee
from where the probability error $\epsilon$ can be determined.
Once $\epsilon$ is found, the phase error $\phaseerror$ can be derived from either Eq.~\eqref{PN-small-approx-pp} or Eq.~\eqref{PN-small-approx-mp}, or from
\be
\frac{P_{2\M}^{++}P_{2\M}^{-+}}{P_{2\M}^{++} + P_{2\M}^{-+}} \approx \sin^2(2\M \xi) \epsilon + O(\epsilon^2) \label{PN-small-approx-phase}
\ee
Alternatively, one can determine $\phaseerror$ from the ratio $P_{4\M}^{++}/P_{2\M}^{++} \approx 4\cos^2(2\M \xi) \epsilon + O(\epsilon^2)$ or from the ratio $P_{2\M}^{-+}/P_{2\M}^{++} \approx \tan^2(\xi)$.

\begin{figure}[tb]
\begin{tabular}{c}
\includegraphics[width=0.90\columnwidth]{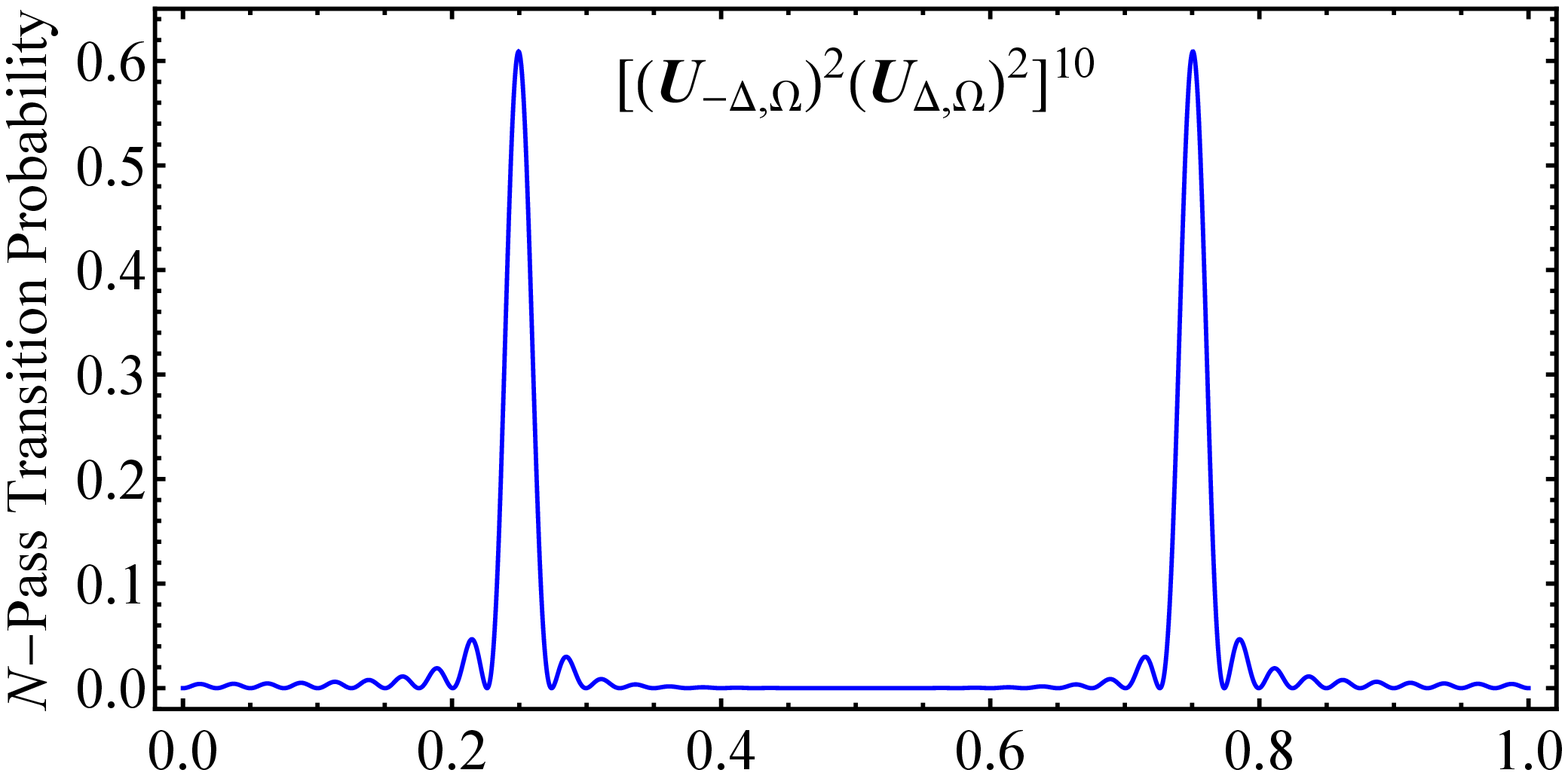} \\
\includegraphics[width=0.90\columnwidth]{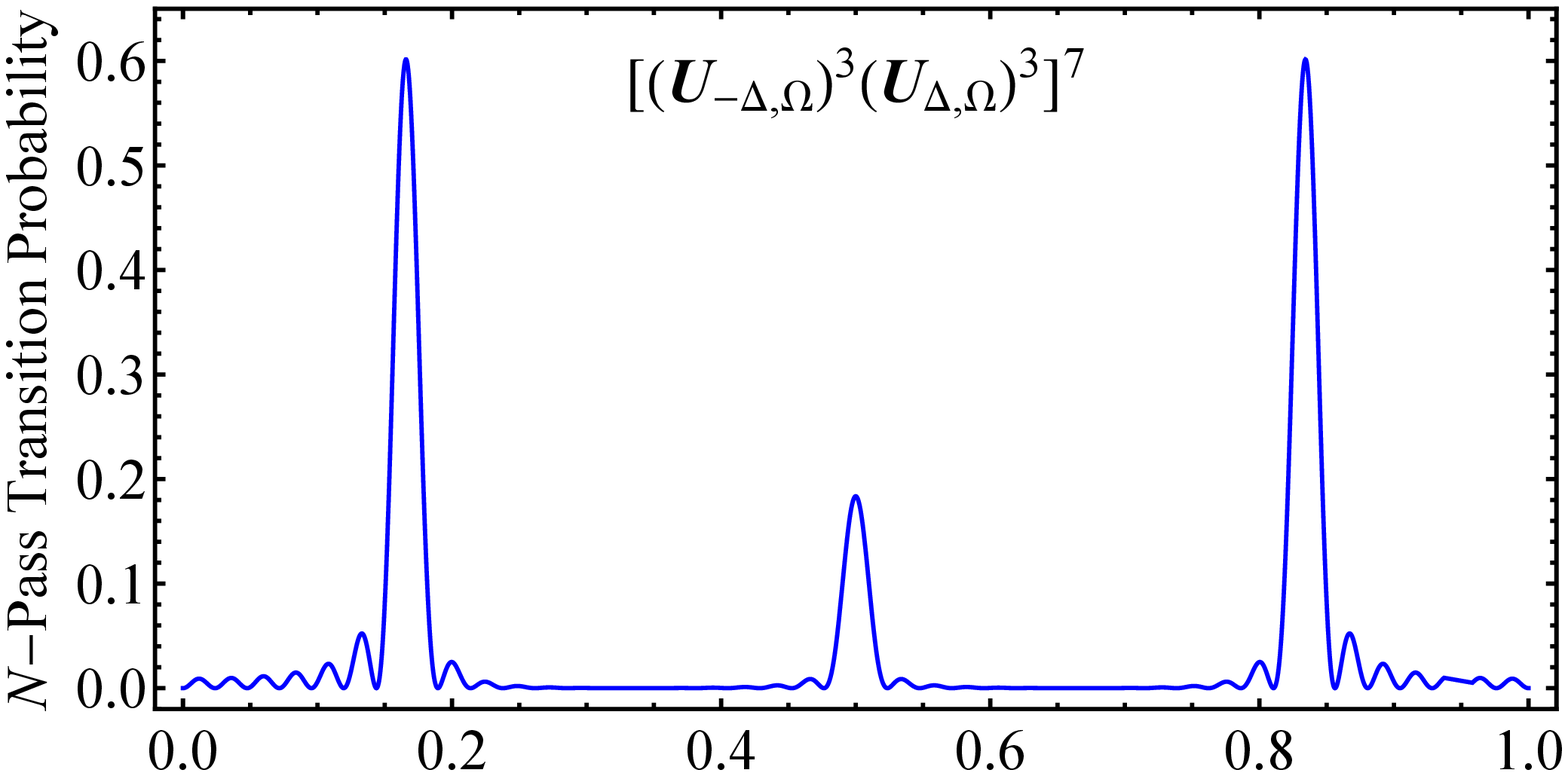} \\
\includegraphics[width=0.90\columnwidth]{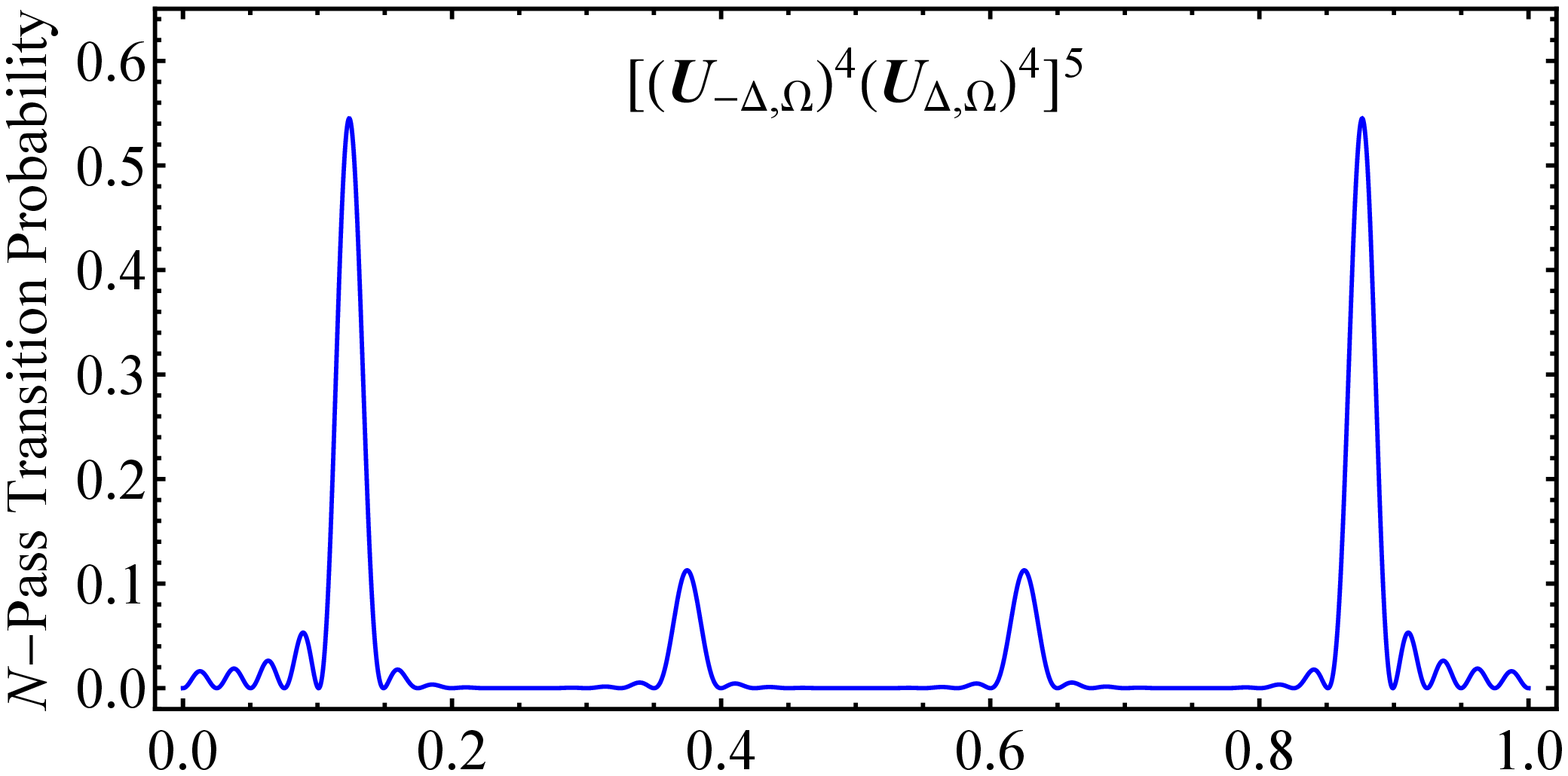} \\
\includegraphics[width=0.90\columnwidth]{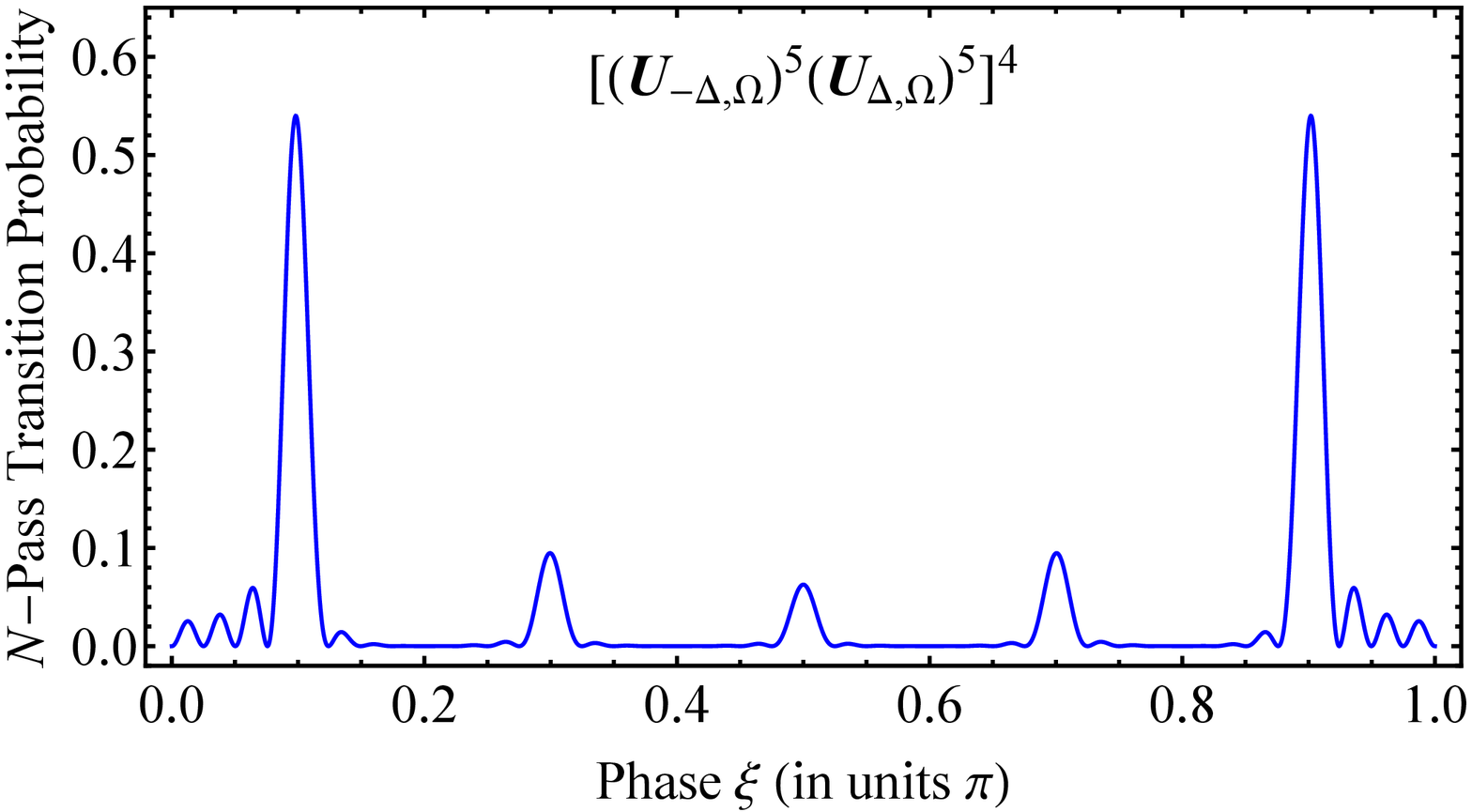}
\end{tabular}
\caption{
Multi-pass transition probabilities stemming from Eq.~\eqref{UN_phasegate} for four different sequences of (errant) phase gates, Eq.~\eqref{phase gate},
 vs the phase $\xi$ of the Cayley-Klein parameter $a$.
The four frames show the transition probabilities after $M$ repeated application of the sequence of $2n$ gates in Eq.~\eqref{UN_phasegate} for a single-pass transition probability $p=0.001$.
The peaks are located at phases $\xi = (2k+1)\alpha$ $(k=0,1,2,\ldots)$, with $\alpha=\pi/(2n)$.
The peak profiles are approximately described by Eq.~\eqref{PN_phasegate}.
Note that only the range $\xi \in [0,\pi]$ is shown because the pattern repeats itself in $[\pi,2\pi]$.
}
\label{fig:phasegate}
\end{figure}

For several important phase gates, there is another, simpler approach to their characterization.
For $\alpha= \pi/(2n)$ one can use the sequences
\be\label{UN_phasegate}
[(\U_{-\Omega,\Delta})^n (\U_{\Omega,\Delta})^n]^M.
\ee
In other words, the original gate $\U_{\Omega,\Delta}$ is repeatedly applied $n$ times, followed by $n$ repetitions of the gate $\U_{-\Omega,\Delta}$ obtained from the original gate by flipping the sign of the Rabi frequency $\Omega$.
This sequence of $2n$ gates in repeated as a whole $M$ times, meaning $N = 2nM$ gates in total.
In fact, the lower frame of Fig.~\ref{fig:small} already shows the transition probability for the case of $n=1$ and $M=10$, which is dominated by peaks at phase values $\xi=\pi/2$ and $3\pi/2$.
The excitation profiles for phases $\alpha = \pi/4,\pi/6,\pi/8,\pi/10$, obtained for $n=2,3,4,5$, are plotted in Fig.~\ref{fig:phasegate}.
These profiles are dominated by sharp peaks at phases $\alpha$, $3\alpha$, $5\alpha$, etc.
Of special importance for quantum information are the first peaks at $\pi/4$ and $\pi/8$ in the cases of $n=2$ and $n=4$ because these phase gates are parts of the basic set of quantum gates.

In the vicinity of $\alpha = \pi/(2n)$ the multi-pass transition probability behaves as
\be\label{PN_phasegate}
P_{N} \approx 4M^2 p \frac{ 1 - 2 \gamma \cot(\alpha) }{\sin^2(\alpha)} .
\ee
This estimate remains valid in the vicinity of the other peaks at $\xi=(2k+1)\alpha/(2n)$ $(k=1,2, \ldots)$ upon the replacement $\alpha \to (2k+1) \alpha$ in Eq.~\eqref{PN_phasegate}.
The single-pass transition probability $p$ can be found by the methods described above, and the numbers $M$ and $n$ are known.
Therefore, this estimate allows one to determine the phase error $\gamma$ by measuring $P_N$.

It should be noted that other profiles with sharp peaks at particular values of the phase $\xi$ can be obtained by sequences similar to Eq.~\eqref{UN_phasegate} by replacing the propagators there by other propagators from Eqs.~\eqref{2ss-UU}.

\black

\section{Summary and conclusions\label{Sec:conclusions}}

\red

In this paper, the relations between the parameters of the single-pass and $N$-pass qubit propagators have been used to assess the coherent amplification of the errors of various single qubit gates due to repeated application of the corresponding gate.
This allows one to determine tiny gate errors by amplifying them to sufficiently high values at which they can be measured with high accuracy by standard quantum process tomography.
Contrary to other methods for error characterization, here no additional errors are introduced besides the error of the considered gate.

In several important special cases the relation between the single-pass transition probability $p$ and the $N$-pass transition probability $\PN$ allows one to unambiguously determine $p$ from $\PN$.
In the most general case, the detrimental multi-pass interference due to the concomitant dynamical phases in the single-pass propagator can be eliminated by taking appropriate sums and ratios of probabilities.
These probabilities are linked to sequences of pairs of interleaved gates, consisting of the original gate and another gate obtained from it by flipping the sign of the Rabi frequency and/or the detuning.
(An important assumption is that the this sign flip, which does not change the gate probabilities, can be performed with very high accuracy.)

%

It is important to note that the quadratic enhancement of the gate error, as suggested earlier for coherent accumulation of errors using heuristic arguments, applies only to some special cases.
In general the dependence of the populations on the gate error is much more involved.
In particular, for a large single-pass transition probability,  $p=1-\epsilon$, the quantum-mechanical probability error increases due to quantum interference much faster in sequential processes than the classical probability: as $N^2\epsilon$ in the quantum case compared to  $N\epsilon$ in the classical case.
%
The same quadratic enhancement of the error occurs in other special cases when the Hamiltonian possesses certain symmetry.
For probability $p=\frac12-\epsilon$, the multi-pass error amplification is linear or quadratic depending on the number of passes $N$: $\propto N\epsilon$ for odd $N$ and $\propto N^2\epsilon^2$ for even $N$.
For small single-pass transition probability, $p\approx \epsilon$, which is important in cross-talk characterization and phase gates, the relation to the $N$-pass transition probability is more complicated because in the general case the latter depends on the dynamical phases of the single-pass propagator.


The relations between the single-pass and multi-pass propagators and probabilities discussed here are a viable alternative of the established randomized benchmarking and other methods for characterizing small gate errors.
One advantage of the present approach is that it can be used for Clifford and non-Clifford gates as well, as it is the case for the quantum phase gate considered in Sec.~\ref{Sec:phase_gate}.
Another advantage is that by repeating the same gate multiple times no new errors are introduced and the only error present in the multi-pass propagator is the sought gate error.
Moreover, in a number of cases the error amplification increases quadratically with the number of passes, which makes the present method much faster than previous methods.
It is also important that in the same time the present method does not depend on SPAM (state preparation and measurement) errors because the amplification of the error to values $O(1)$ renders SPAM errors negligible.

\black

\acknowledgments

This work is supported by the 
 Bulgarian Science Fund Grant DO02/3 (ERyQSenS).


\end{document}